\documentclass[dvips, final]{article}

\usepackage{icrc2011}

\title{A new Monte Carlo Generator for Ultra-High Energy Cosmic Rays from the 
Local and Distant Universe}

\newcommand{\etal}{\MakeLowercase{\textit{et al. }}} 
\shorttitle{Dolag \etal UHECR Monte Carlo}

\authors{Klaus Dolag $^{1}$, Martin Erdmann$^{2}$, Gero M\"uller$^{2}$, David Walz$^{2}$, Tobias Winchen$^{2}$}
\afiliations{$^1$ MPI f\"ur Astrophysik, Garching, Germany \\ $^2$ III. Physikalisches Institut A,
RWTH Aachen University, Germany}
\email{tobias.winchen@physik.rwth-aachen.de}

\abstract{
For the understanding of the origin and propagation of ultra-high energy
cosmic rays (UHECR) we developed a new approach to simulating UHECRs
from an arbitrary number of sources based on Monte Carlo technique.  The
method consists of a combination of three steps. For distant sources we
apply commonly accepted parameterizations to calculate the contribution
to the observed cosmic ray flux.  For sources of the local universe we
use forward tracking through realistic matter distributions and magnetic
fields resulting from explicit simulations of large-scale structure
formation.
From the calculations and the forward tracking we generate maps of the
probability to observe a particle with a given energy from a discrete
direction.  To account for deflections in the galactic field, these
probability maps are transformed by matrices calculated from backtracking of
antiparticles through field parameterizations.  Based on the combined
probability maps, Monte Carlo production of individual UHECR data is
performed which are then used in comparisons with UHECR measurements. 
The simulated UHECR data serves for optimizing the
analysis techniques used in UHECR measurements as well as for constraining 
the parameter space of the underlying source and magnetic field
models.
}
\keywords{UHECR Propagation, Monte Carlo simulations}

\begin{document}
\maketitle

\section{Introduction}

The comparison of observations with model predictions by means of Monte
Carlo generated data can be a powerful tool to improve the knowledge on
the origin and propagation of UHECRs. Currently there are basically two
approaches to this challenge.

In the first approach UHECR data are generated by a full
forward simulation of the UHECR propagation from their sources to the
observer, accounting for energy losses and deflections in magnetic fields
at every point of the trajectory. The low rate of trajectories hitting
the observer renders the generation of large datasets by this technique a
computationally intensive task. It becomes more and more costly as the
source distances increase. Furthermore energy losses due to the
expansion of the universe and the corresponding increase of the photon
density are difficult to incorporate efficiently.

A second approach uses parametrizations for the energy losses and
deflections in magnetic fields and is hence much less computationally
intensive than the full forward simulation.  These parametrized
simulations however cannot easily describe structured extragalactic
magnetic fields as expected from large-scale structure simulations.
Furthermore the commonly used parametrizations are the result of an
averaging process over long propagation distances and are thus not
reliable on short propagation distances. However, energy losses due to the
expansion of the universe and the changes in the photon fields can
be easily included. 

In the following we present a combination of both approaches for the
propagation of protons.
For the UHECR flux from sources up to a distance of 110 Mpc, in the
following called the \emph{local universe}, we track individual
particles using matter distributions and magnetic fields
from constrained simulations of large-scale structure
formation. The size of this simulation regime is determined by the size
of the structure simulations at hand. Furthermore a current proper
distance of 110 Mpc correspond to a red shift of $z = 0.027$ or a
scaling factor of $R = 0.97$ respectively. Neglecting the increased
energy losses and only accounting for cosmic time dilation the
luminosity of sources in the maximum distance is overestimated by at least 3\%
depending on the magnetic field strength.

\begin{figure*}[th]
	\centering
	\includegraphics[width=3.2in]{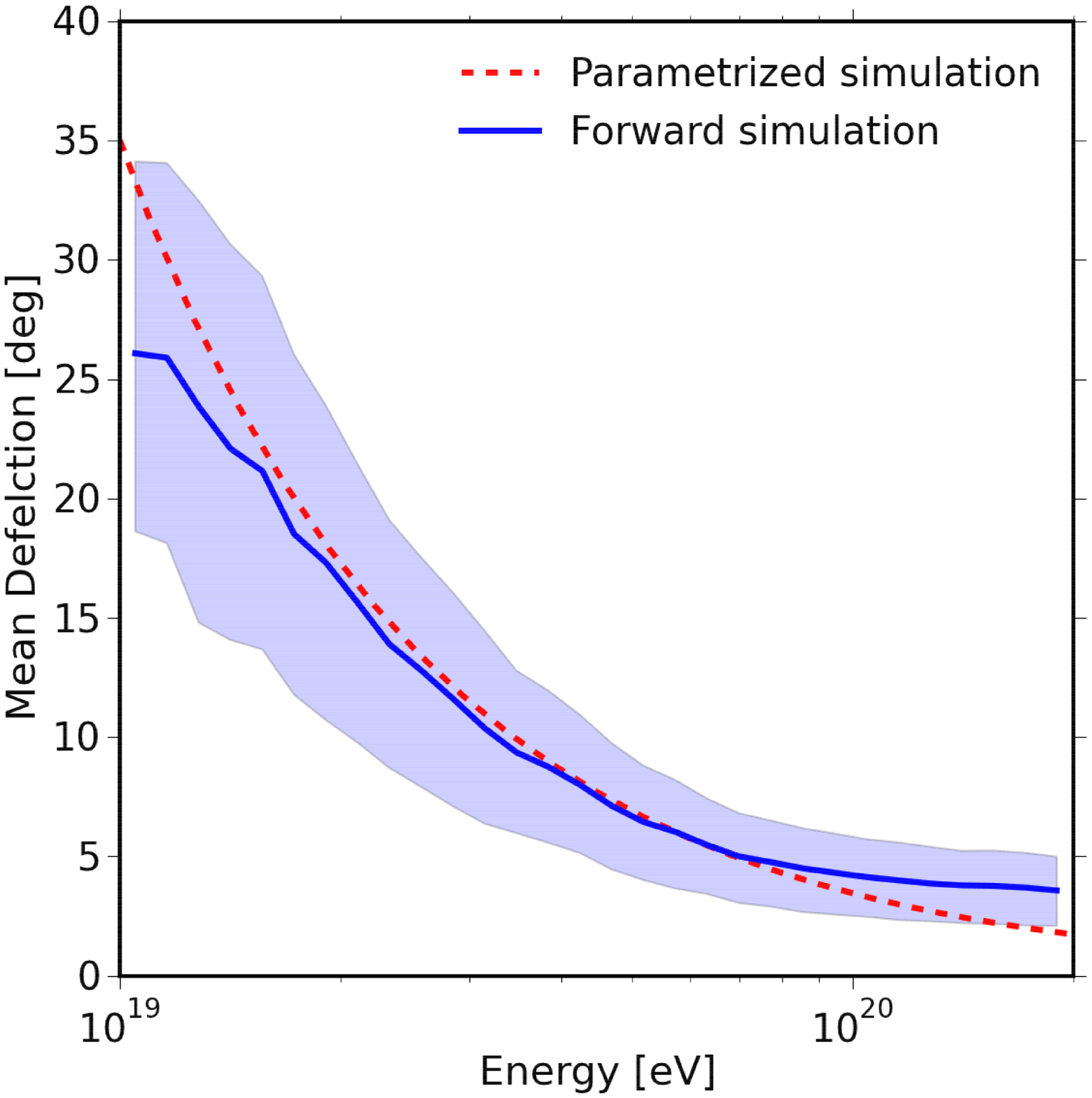}
	\includegraphics[width=3.2in]{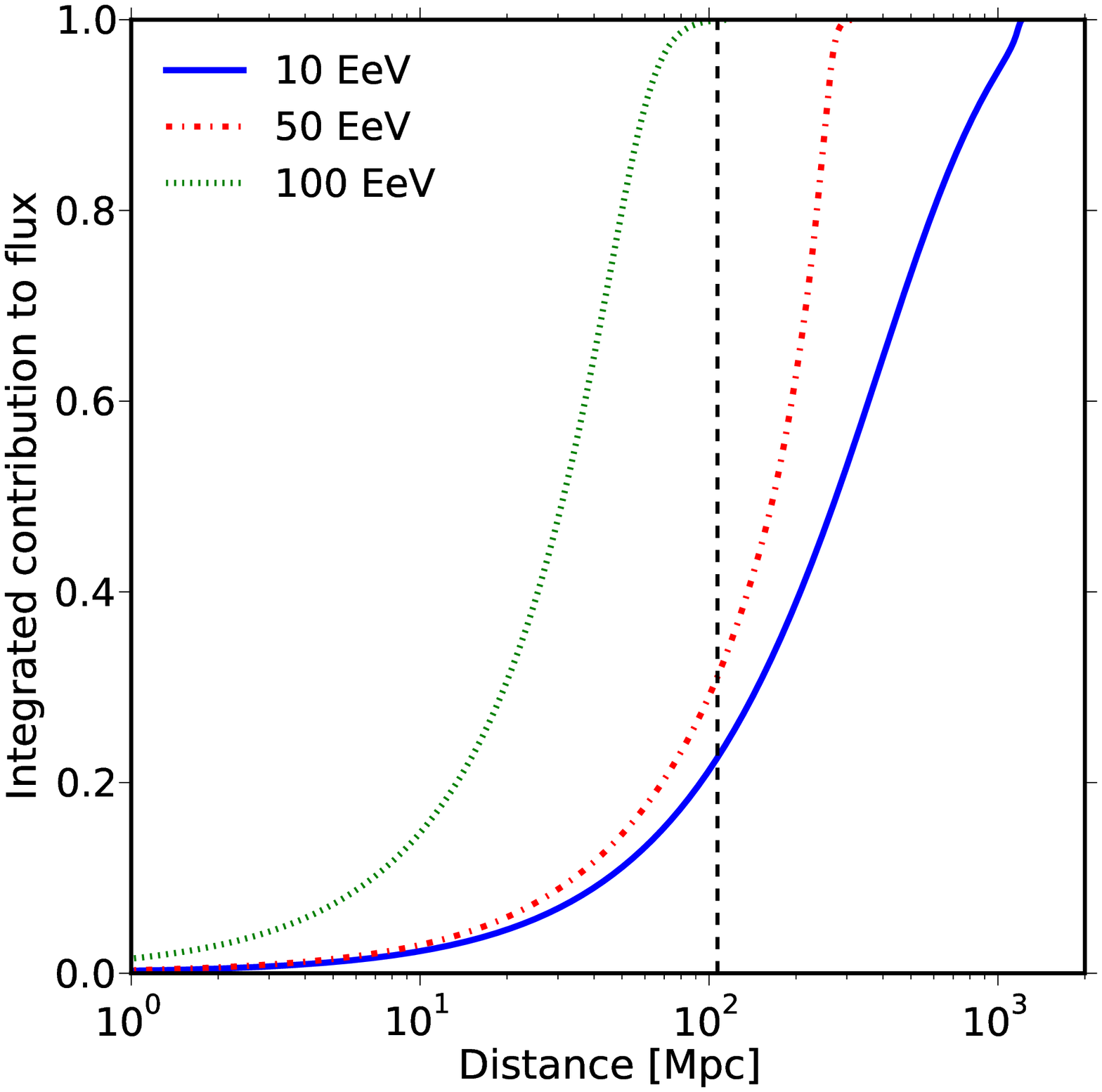}
	\label{fig:matching}
	\caption{Average deflection strength of the forward simulation of
	UHECRs from sources in 107 Mpc distance in an
	exemplary local	universe and best fit of simulations with PARSEC (left panel). Integrated
	contribution from sources up to distance $d$ to the flux for different
	energies (right panel). The vertical dashed line denotes the merging
	point of the two simulation regimes.}
\end{figure*}

\section{Forward simulation}
The magnetic fields and source distributions used in the forward simulation are
the result from simulations of $\Lambda$CDM large scale structure
formation \cite{Dolag2005}. The simulations have been constrained using
data from the IRAS 1.2Jy Redshift
Survey \cite{Fisher1995} to reproduce the observed distribution of luminous matter.  
These simulations have been combined with simulations of the
magneto hydrodynamic evolution of a primordial seed field.
The resulting magnetic fields are shown to be compatible
with observations by means of synthetic Faraday
measurements.

In our simulations UHECRs are injected at point sources with a
distribution following the distribution of matter in the
large-scale structure simulation. The trajectories of the UHECRs from
the source to a spherical observer in the center of the simulation
region are calculated using a modified version of the CRPropa
software \cite{Armengaud2007,Erdmann2011a}.

The size of the observer dominates the detection rate of UHECRs in the
forward simulation. Larger observers however introduce systematic errors
in the deflection angles due to the geometric distortion and reduced
propagation path lengths. To account for theses effects and also ensure
a high performance we adapt the observer size to the distance of the
source such that the systematic error on the deflection remains below 2
deg.  

The simulations of the large-scale structure are performed in the
smoothed particle hydrodynamics formalism.
To avoid computationally expensive calculations we pre-calculate  
the magnetic field from the smoothed particle data on a regular cubic grid with 100
kpc distance between nodes. The magnetic field at arbitrary positions
is then calculated using trilinear interpolation between the values at
the nodes of a cell.
The grid spacing of
100 kpc however results in a data size of 150 GB for the complete
grid. To avoid swapping due to the limited memory of typical desktop
PCs, the grid is divided into cubes with an edge length of 20 Mpc.
The resulting memory load of ca. 1 GB for the magnetic field can
be easily handled by
current desktop PCs.
All UHECRs in one box are first propagated to the edges of the cube.
Second the next cube is loaded into memory and all corresponding UHECRs are
processed. The processing of the individual cubes is optionally parallelized on
several CPUs. 
To further improve the performance, UHECRs with a propagation distance
larger than 1.5 times the linear distance between observer and source
are dropped. Only about 1 per mille of the UHECRs are lost owing to
this cut \cite{Erdmann2011a}.

\begin{figure*}[th]
	\centering
		\includegraphics[width=.9\textwidth]{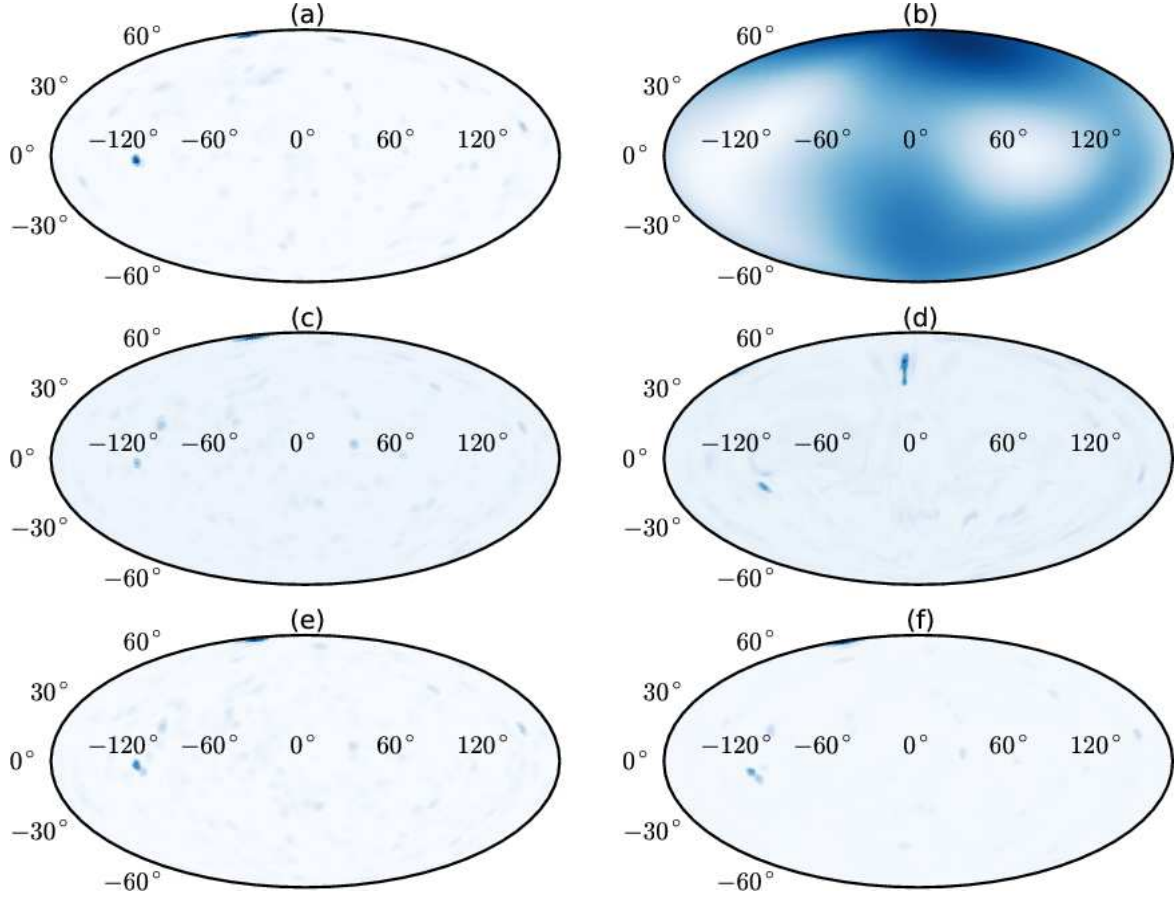}
	\caption{Hammer projected probability density maps in
	galactic coordinates for an exemplary
	realization with a source density of $10^{-3.5} \mathrm{Mpc}^{-3}$. 
	Blue indicates higher, white lower probability on an arbitrary scale.
	\textbf{(a)}  Forward simulation for 10 EeV. \textbf{(b)} Parametrized simulation for 10 EeV. 
	\textbf{(c)} Weighted combination of map a) and b). \textbf{(d)} Application
	of a BSS\_S \cite{Harari1999} type galactic magnetic field to map c).
	\textbf{(e)} Probability map from the forward simulation for 50 EeV. 
	\textbf{(f)}  Application of a BSS\_S type galactic magnetic field to
	map e).}
	\label{fig:ExemplarySkyMaps}
\end{figure*}

\section{Parametrized simulation}
For the propagation in the distant universe we use the PARSEC software
\cite{Erdmann2011}, which uses parametrizations for the energy loss and
deflections in turbulent fields to calculate a map of the probability
to observe a particle in a discrete direction (pixel) in a given energy
bin for extragalactic propagation. To account for deflections in
galactic magnetic fields, these probability maps are then
transformed using
pre-calculated matrices obtained from backtracking of UHECRs through
models of the galactic field.  Energy losses are modelled within continuous energy
loss approximation implementing the attenuation length from
\cite{Protheroe1996} in the extragalactic propagation and neglected in our 
galaxy.

Assuming source spectra following a power law with
spectral index $\gamma$, the probability to observe a particle in pixel
$j$ with energy $E_i$ in the range $E_{l,i} < E_i
\leq E_{u,i}$ is 
\begin{equation}
	p_j^i = \Gamma_i \sum_k \frac{L_k}{d_k^2
	(1+\hat{z}_{k})} (\hat{E}_{u,i}^{\gamma+1} - \hat{E}_{l,i}^{\gamma+1})
	P_{EGMF}(\alpha_{j,k}).
	\label{probabilityEquation}
\end{equation}
Here $\hat{z}_k$ is the redshift of the source at time of emission of the particles at
source $k$ in current proper distance $d_k$ with luminosity $L_k$.
$\hat{E}_{i,l}$ and $\hat{E}_{i,u}$ denote the energies at the source contributing to
the observed energies in the range $E_{l,i} - E_{u,i}$ and include energy
losses from photon interaction and expansion of the universe. The factor
$P_{EGMF}(\alpha_{j,k})$ accounts for the deflection of the UHECR in extragalactic
fields  with $\alpha_{j,k}$ being the angle between the direction of
pixel $j$ and the direction of source $k$.

To describe the deflection in the extragalactic magnetic fields we use a
Fisher distribution \cite{Fisher1953} to describe the angular distribution of the
UHECRs around the source. The Fisher distribution is the
normal-distribution on a sphere and has the functional form 
\begin{equation}
	f(\kappa, \alpha) = \frac{\kappa}{4\pi \sinh{\kappa}}
	e^{\cos{\alpha}}
	\label{iFisherDistribution}
\end{equation}
with concentration parameter $\kappa$ indicating the width of the
distribution and $\alpha$ the angular distance from the center of the
distribution. For small concentration parameters $\kappa$ the Fisher
distribution converges to the isotropic distribution. For large $\kappa$
the Fisher distribution converges to a Rayleigh distribution with width
$\sigma$ and $\kappa = 1/\sigma^2$. 

We use $\kappa = 1/\sigma^2$ with
\begin{equation}
	\sigma \propto \sqrt{D_k \Lambda} \frac{B}{E}
	\label{sigmaField}
\end{equation} 
taken from \cite{Achterberg1998} to describe the deflection power in magnetic
fields. Here
$D_k$ is the distance of source $k$, $
\Lambda$ the coherence length, and  $B$ the strength 
of the magnetic field. $E$ denotes the energy of the cosmic ray. The elongation
of the propagation path due to deflection in the magnetic fields is parametrized based
on the work of Achterberg et al. \cite{Achterberg1998}, but has been modified to include energy
losses in the parametrization \cite{Erdmann2011}.

\section{Combination of simulation regimes}
For a consistent model, the deflection strength of the parametrized
simulation has to be scaled to match the mean deflection in the forward
propagation. The left panel of figure \ref{fig:matching} shows the mean deflection
resulting from the forward simulation of 5000 UHECRs from
isotropically distributed starting points in 107 Mpc distance in the
local universe and the best fit of the mean deflection strength with
PARSEC. The best
fit is achieved for $B \sqrt{\Lambda} = 0.94
~\mathrm{nG}\sqrt{\mathrm{Mpc}}$ in equation \ref{sigmaField}.

Both simulation regimes are combined by means of summing up maps of the
probability to observe a particle in a given energy range from a
discrete direction. While the probability maps of the contribution from
the distant universe are calculated explicitly with PARSEC, the
probability maps from the local universe are created from the forward
tracking data.  The maps are weighted so that the total contribution
from the local universe matches the expected contribution from
continuous sources in the same volume for the lowest energy bin.  This
weight is calculated using PARSEC with the matched strength for the
magnetic field. The integrated relative contributions over the distance
are shown in the right panel of figure \ref{fig:matching} for three
different energies .

To account for deflections in the galactic field, the resulting
probability maps are transformed using the matrix technique provided by
PARSEC. From the final probability maps simulated data sets with large
numbers of UHECRs are then generated with low computational effort. 

\section{Results and Conclusion}
\begin{figure}[t]
  \centering
  \includegraphics[width=.48\textwidth]{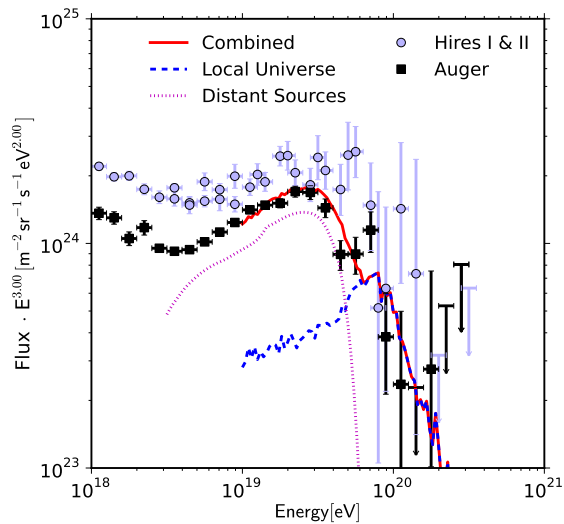}
	\caption{Energy spectrum resulting from the exemplary simulation in
	comparison with observations from the HiRES experiment
	\cite{HiRes2008} and the Pierre-Auger Observatory \cite{PAO2010a}.}
	\label{fig:spectrum}
\end{figure}

In figure \ref{fig:ExemplarySkyMaps} we show exemplary probability maps
from the forward propagation in the local universe and the parametrized
simulation in the distant universe together with the combined maps at two
energies. The source density in this exemplary simulation is $\rho = 10^{-3.5}$
Mpc$^{-3}$, resulting in a total number of 1763 sources in the local universe
and more than 4.4 million sources up to the maximum simulation distance
of 1500 Mpc. The source spectra have been set to $\gamma = -2.7$.
For an energy of 10 EeV the ratio of the flux from the distant
universe to the local universe $N_{near} / N_{tot} = 0.23$. For energies
above 100 EeV all particles originate from the local universe.

In figure \ref{fig:spectrum} we show the resulting energy spectrum from the
exemplary realization calculated using 100 energy bins between $10^{18.5}$
eV and $10^{20}$ eV. For comparison the observed energy spectra of
the HiRes experiment \cite{HiRes2008} and the Pierre-Auger Observatory
\cite{PAO2010a} are also shown. The
simulated spectrum is normalized to match the observation of the Pierre
Auger Observatory at 10 EeV.

With the technique presented in this contribution we have developed a promising
mechanism for the extensive production of Monte Carlo simulations of
UHECRs suitable for studies of anisotropy and cosmic
magnetic fields. The simulation includes effects of structured extragalactic and
galactic magnetic fields, as well as energy losses and arbitrary source
configurations. 

%
%

\clearpage

\end{document}